%% file: main.tex
\documentclass[%
    reprint,
    superscriptaddress,
    showpacs,
    amsmath,
    amssymb,
    aps,
    prd,
    nofootinbib,
]{revtex4-2}

\input{definitions}

\begin{document}


\title{Test of lepton flavor universality with \boldmath{$\bar{B} \rightarrow D^{(*)} \tau^{-} \bar{\nu}_{\tau}$} \\ and \boldmath{$\bar{B} \rightarrow D^{(*)} \ell^{-} \bar{\nu}_{\ell}$} decays at Belle~II}
\input{pub101-orcid}

\begin{abstract}
We test lepton flavor universality with a measurement of the branching-fraction ratios \mbox{$R(D^{(*)}) \equiv \mathcal{B}(\bar{B} \rightarrow D^{(*)} \tau^{-} \bar{\nu}_{\tau})/\mathcal{B}(\bar{B} \rightarrow D^{(*)} \ell^{-} \bar{\nu}_{\ell})$}, where $\ell$ denotes an electron or muon. The analysis uses $387\times 10^6$ $\Upsilon(\mathrm{4S})$ decays collected with the Belle~II detector in energy-asymmetric $e^+e^-$ collisions. One $B$ meson is fully reconstructed in a hadronic decay mode, while the other is reconstructed either in $\bar{B}\rightarrow D^{(*)}\tau^{-}\bar{\nu}_{\tau}$, with $\tau^- \rightarrow \ell^- \bar{\nu}_{\ell}\nu_{\tau}$, or in $\bar{B}\rightarrow D^{(*)}\ell^{-}\bar{\nu}_{\ell}$. We extract the signal from the distributions of the residual calorimeter energy and the squared mass of the undetected particles, obtaining $R(D^{*}) = 0.242 \pm0.019(\mathrm{stat}) \pm0.016(\mathrm{syst})$ and $R(D) = 0.439 \pm 0.055(\mathrm{stat}) \pm 0.046(\mathrm{syst})$. These results are consistent with both the standard model predictions and previous measurements, and constitute the most precise determination of $R(D^{(*)})$ with hadronic tagging. 

\end{abstract}

\maketitle

In the standard model (SM), the electroweak interaction couples equally to the three lepton families, a property known as lepton flavor universality (LFU)~\cite{LEP_LU, ATLAS_LU, CMS_LU}. Any deviation from LFU would signal physics beyond the SM~\cite{London_2022, LFU_review2021}. 
 Semileptonic $B$ decays provide a sensitive test as several theoretical and experimental uncertainties cancel in the ratios of branching fractions across different lepton flavors. The ratios  
\begin{equation}
R(D^{(*)}) \equiv \frac{\mathcal{B}(\bar{B} \rightarrow D^{(*)} \tau^{-} \bar{\nu}_{\tau})}{\mathcal{B} (\bar{B} \rightarrow D^{(*)} \ell^{-} \bar{\nu}_{\ell})}\,,
\end{equation}
probe LFU between the third lepton family ($\tau$) and the first two ($\ell = e, \mu$). 
Here, $D^{(*)}$ denotes either $D^{(*)+}$ or $D^{(*)0}$, and $\bar B$ denotes either $B^-$ or $\bar B^0$. 
The world average~\cite{HFLAV:2024} of existing measurements~\cite{BabarDtaunu1, BabarDtaunu2, BelleDtaunu1, BelleDtaunu3,BelleDtaunu4, LHCbDtaunu, LHCbDtaunu2, LHCb2023, LHCb2023_2,Belle2RDstar,Belle2RDRDstar} shows a persistent excess relative to SM predictions~\cite{SM_RDRDstar1, SM_RDRDstar2, SM_RDRDstar3, BLPRXP, SM_RDRDstar5_updated, SM_RDRDstar6, SM_RDRDstar7, SM_RDRDstar8, SM_RDRDstar9}, suggesting either LFU violation or underestimated uncertainties.
The inclusion of charge-conjugated decay modes is implied throughout. 

In this Letter, we present a simultaneous measurement of $R(D)$ and $R(D^{*})$ using $387\times 10^6$ $\Upsilon(\mathrm{4S})$ decays collected by the Belle II experiment between 2019 and 2022. 
One $B$ meson, $B_{\mathrm{tag}}$, is fully reconstructed in a hadronic decay mode (hadronic tagging), while the signal, $B_{\mathrm{sig}}$, is reconstructed in either $\bar{B}\rightarrow D^{(*)} \tau^{-}\bar{\nu}_{\tau}$, with $\tau^- \rightarrow \ell^- \bar{\nu}_{\ell}\nu_{\tau}$, or $\bar{B}\rightarrow D^{(*)} \ell^{-}\bar{\nu}_{\ell}$. This result improves our earlier $R(D^*)$ measurement with half the data~\cite{Belle2RDstar},  provides the first hadronic-tagged measurement of $R(D)$ at Belle~II, and enhances sensitivity to $R(D^*)$ by including additional $D^{(*)}$ decay channels and feed-down from unreconstructed $D^*$ decays into the $D$ samples used for the $R(D)$ determination. 

The ratios are extracted with a two-dimensional fit to the distribution of the residual calorimeter energy $E_{\mathrm{ECL}}$ and the missing-mass squared $M^2_{\mathrm{miss}}$. The residual energy is the sum of energy deposits in the calorimeter not associated with the reconstructed $B_{\mathrm{sig}}$ and $B_{\mathrm{tag}}$. 
The missing-mass squared is defined as the invariant-mass squared of the undetected particles, computed from the four-momentum remaining after subtracting the four-momenta of $B_{\mathrm{tag}}$ and the $D^{(*)}\ell$ system from the $e^+e^-$ four-momentum. The $B_{\mathrm{tag}}$ energy is constrained to half of the beam energy in the center-of-mass frame.

In signal events, where only neutrinos are missing, $M^2_{\mathrm{miss}}$ peaks near zero for $\bar{B} \rightarrow D^{(*)} \ell^{-} \bar{\nu}_{\ell}$ and at higher values for $\bar{B} \rightarrow D^{(*)} \tau^{-} \bar{\nu}_{\tau}$, while $E_{\mathrm{ECL}}$ peaks near zero for both decays. Backgrounds with additional undetected particles yield broader distributions with larger values of $M^2_{\mathrm{miss}}$ and $E_{\mathrm{ECL}}$. 

The analysis strategy is developed using simulated samples and signal-depleted data control regions. The signal region is examined only after all procedures are finalized.


The Belle II detector~\cite{belle2tdr} at the SuperKEKB asymmetric-energy $e^+e^-$ collider~\cite{SuperKEKB} consists of a nearly hermetic solenoidal magnetic spectrometer with a silicon vertex detector comprising pixel and strip sensors, a central drift chamber for charged-particle tracking, and surrounding particle-identification, electromagnetic-calorimeter, and muon-detection subsystems. 

We generate $e^+e^- \rightarrow \Upsilon(\mathrm{4S}) \rightarrow B\bar{B}$ and continuum $e^+e^- \rightarrow q\bar{q}$ ($q=u,d,s,c$) samples with fractions fixed according to known cross sections and a total size about four times that of the data. 
Particle production and decay are simulated with \texttt{EVTGEN}~\cite{Lange:2001uf}, \texttt{PYTHIA}~\cite{pythia8.3}, and \texttt{KKMC}~\cite{kkmc}, using \texttt{PHOTOS}~\cite{PHOTOS,PHOTOS_new} for final-state radiation and \texttt{GEANT4}~\cite{geant4} for detector response, which is validated and corrected using control samples in data. Simulated beam backgrounds are overlaid~\cite{beambkg}, and all samples are processed with the Belle~II framework \texttt{basf2}~\cite{basf2,basf2-zenodo} as for data.

In simulation, the decays $\bar{B} \rightarrow D^{(*)} \ell^{-} \bar{\nu}_\ell$ and $\bar{B} \rightarrow D^{**} \ell^{-} \bar{\nu}_\ell$ are modelled with the BLPRXP~\cite{BLPRXP} and BLR~\cite{BLR1, BLR2} form factors, respectively. Here, $D^{**}$ denotes the orbitally excited states $D^{*}_{0}$, $D_{1}^{\prime}$, $D_{1}$, and $D^{*}_{2}$, which decay into $D^{(*)}\pi(\pi)$.  
The branching fractions of these semileptonic decays use the latest averages from Ref.~\cite{HFLAV:2024} and the calculation follows the assumptions described in Ref.~\cite{Belle2RDstar} for $\bar{B} \rightarrow D^{**} \ell^{-} \bar{\nu}_\ell$. 
While  $\bar{B} \rightarrow D^{**}\ell^{-} \bar{\nu}_\ell$ decays saturate the inclusive $\bar{B} \rightarrow D^{(*)}\pi \ell^{-} \bar{\nu}_\ell$  rate, a nonresonant component is needed for $\bar{B} \rightarrow D^{(*)}\pi\pi \ell^{-} \bar{\nu}_\ell$ and its branching fraction is calculated based on the same assumptions as in Ref.~\cite{BaBarDstpipi} and with updated input measurements~\cite{HFLAV:2024}. In addition, to account for the difference between the total inclusive  semileptonic rate and the sum of the exclusive rates, we include $\bar{B} \rightarrow  D^{(*)}\eta\ell^{-}\bar{\nu}_\ell$ decays (gap modes).
Both nonresonant $\bar{B} \rightarrow D^{(*)}\pi\pi \ell^{-} \bar{\nu}_\ell$ and $\bar{B} \rightarrow  D^{(*)}\eta\ell^{-}\bar{\nu}_\ell$ are simulated as equal mixtures of $D_{0}^{*}$ and $D_{1}^{\prime}$ resonances with BLR form factors. For the semitauonic decays $\bar{B} \rightarrow D^{(*)} \tau^{-} \bar{\nu}_\tau$ and $\bar{B} \rightarrow D^{**} \tau^{-} \bar{\nu}_\tau$, including gap modes, we use the same models of the light-lepton counterparts, with branching fractions  derived assuming LFU~\cite{DssAve1, DssAve2, DssAve3}.


To identify signal events, we reconstruct the $B_{\mathrm{tag}}$ meson in fully hadronic decay modes~\cite{fei}. The $B_{\mathrm{tag}}$ selection follows Ref.~\cite{Belle2RDstar}, with the additional requirement $\cos\alpha < 0.9$ to suppress continuum background. Here, $\alpha$ is the angle between the $B_{\mathrm{tag}}$ thrust axis~\cite{Thrust-axis, Thrust} and that of the rest of the event. 
The $B_{\mathrm{tag}}$ reconstruction has an average efficiency of 0.30\% and a purity of 29\%.

The $B_{\mathrm{sig}}$ candidate is obtained from a $D^{(*)}\ell$ combination reconstructed from the remaining tracks and calorimeter-energy clusters. 
The $D^{*}$ mesons are reconstructed in the decay channels $D^{*+} \rightarrow D^0 \pi^+, D^+ \pi^0$ and $D^{*0} \rightarrow D^0 \pi^0, D^0 \gamma$, with the $D^0\gamma$ channel being an addition with respect to Ref.~\cite{Belle2RDstar}. 
To maximize signal sensitivity, $D$ mesons are reconstructed in seven modes covering 36\% of the $D^0$ and 29\% of the $D^+$ total widths. The $D^0$ channels are:  $ K^- \pi^+$, $ K^0_{S} \pi^0$, $ K^- K^+$, $ K^0_S \pi^+ \pi^- $, $K^- \pi^+ \pi^0$, $ K^- \pi^+ \pi^- \pi^+$ and $ K^0_{S} \pi^+ \pi^- \pi^0$. The $D^+$ channels are: $ K^0_S \pi^+$, $K^- \pi^+ \pi^+$, $ K^- K^+ \pi^+$, $ K^0_S \pi^+ \pi^0$, $ K^- \pi^+ \pi^+ \pi^0$, and $K^0_S K^+$. The last three are new with respect to Ref.~\cite{Belle2RDstar}.

Tracks are required to originate near the interaction point, with a distance of closest approach within 4.0\,cm (2.0\,cm) along (transverse to) the $z$~axis. 
Charged particles must have a transverse momentum above 0.1\,GeV/$c$, except for low-momentum pions from $D^{*+} \rightarrow D^0 \pi^+$ decays. 
Momenta are calibrated using control samples of $D^{*+}\to D^0(\to K^-\pi^+)\pi^+$ prior to any selection.
Electron (muon) candidates must have momentum above 0.4\,GeV/$c$~(0.7\,GeV/$c$). Neutral kaons are reconstructed in $K^0_S\to\pi^+\pi^-$ decays using the multivariate algorithm of Ref.~\cite{KsFinder}. 

Neutral pions are reconstructed in $\pi^0\to\gamma\gamma$ decays from pairs of photons with minimum energies of 0.08, 0.03, and 0.06\,GeV,  in the forward, barrel, and backward regions, respectively. 
The photon selection follows Ref.~\cite{Belle2RDstar} and the $\pi^0$ invariant-mass is required to be within [0.122, 0.143]\,GeV/$c^2$, retaining 94\% of correctly reconstructed candidates. For low-momentum $\pi^0$ or $\gamma$ from $D^*$ decays, looser  photon criteria are applied, improving the reconstruction efficiency by about 20\%.

For $D^{(*)}$ meson candidates, the reconstructed $D$ mass and the mass difference between $D^{*}$ and $D$, $\Delta M = M_{D^*}-M_{D}$, must be consistent with their known values within 1.2--5.0 times the mass resolution, depending on the decay mode. 

The $\Upsilon(\mathrm{4S})$ candidates are reconstructed from combinations ($B_{\mathrm{sig}}^{+}$,  $B_{\mathrm{tag}}^{-}$),  ($B_{\mathrm{sig}}^{0}$,  $\bar{B}_{\mathrm{tag}}^{0}$), ($B_{\mathrm{sig}}^{0}$,  $B_{\mathrm{tag}}^{0}$), and their charge conjugates. In the ($B_{\mathrm{sig}}^{0}$,  $B_{\mathrm{tag}}^{0}$) combination,  one $B^0$ meson has undergone flavor mixing before decaying.
To suppress background, we reject events with remaining tracks or $\pi^0$ candidates not associated with $\Upsilon(\mathrm{4S})$ candidates. Clusters from beam background or charged-particle deposits are rejected using multivariate algorithms~\cite{photonmva} and minimum cluster-energy requirements of 0.05\,GeV and 0.1\,GeV in the barrel and end-cap regions, respectively.
We enrich the $\bar{B}\rightarrow D^{(*)} \tau^{-}\bar{\nu}_{\tau}$ sample by requiring the squared four-momentum transfer between the $B_{\mathrm{sig}}$ and $D^{(*)}$ four-momenta, $q^{2}$, to exceed 4~GeV$^2/c^2$, where the $B_{\mathrm{sig}}$ four-momentum is derived from that of $B_{\mathrm{tag}}$ using energy-momentum conservation.

After the full selection, multiple $\Upsilon(4S)$ candidates remain per event, with an average multiplicity of 1.26. We retain a single candidate according to the following priority: (i) the highest $B_{\mathrm{tag}}$ probability; (ii) $D^*$ decay-channel order $D^0\pi^+$, $D^+\pi^0$, $D^0\pi^0$, $D^0\gamma$, and, for $D$ decays, $D^0$ before $D^+$; (iii) for $\pi^+$ or $\pi^0$ from $D^*$, the candidate with the best $B_{\mathrm{sig}}$ vertex $\chi^2$ or with a $\pi^0$ mass closest to the nominal value, and, for $\gamma$ from $D^*$, the highest photon energy; (iv) the $D$ mode with the highest branching fraction. If multiple candidates remain, one is chosen at random.

The overall reconstruction and selection efficiencies are calculated from simulation to be in the ranges $(0.62$--$4.77) \times 10^{-5}$ and $(1.66$--$12.5) \times 10^{-5}$ for $\bar{B}\to D^{(*)}\tau^-\bar\nu_\tau$ and $\bar{B}\to D^{(*)}\ell^-\bar\nu_\ell$, respectively, depending on the $D^{(*)}$ decay mode. These are corrected for known data-simulation discrepancies using dedicated control samples. Tracking efficiency is calibrated with $e^+e^- \rightarrow \tau^+\tau^-$ events for charged particles with transverse momentum between 0.2 and 2\,GeV$/c$, and with $\bar{B}^0 \to D^{*+}(\to D^0\pi^+)\pi^-$ decays for lower momentum particles. Charged-particle identification is corrected with samples of $e^+ e^- \to e^+ e^- \ell^+ \ell^-$, $e^+ e^- \to \ell^+ \ell^- (\gamma)$, $J/\psi \to \ell^+\ell^-$, $K^0_S \to \pi^+\pi^-$, and $D^{*+} \to D^0(\to K^-\pi^+)\pi^+$~\cite{B2HadronID}. Photon and $\pi^0$ reconstruction are calibrated with $e^+ e^- \to \mu^+ \mu^- \gamma$, $B^- \to D^{*0}(\to D^0\pi^0)\pi^+$, and $D^{0} \to K^-\pi^+(\pi^0)$. Hadronic tagging is validated with $\bar B\to X\ell^- \bar \nu_{\ell}$ and $\bar B \to D^{(*)}\pi^-$ decays, where $X$ denotes any hadronic system,  and decay-mode-dependent corrections are determined~\cite{FEICalib}. 
The average of all simulation corrections combined ranges from 5\% to 20\%, depending on the $D^{(*)}$ mode, where the largest effect is due to the low-momentum $\pi^0$ correction. 

 The $D^{(*)}$ mass resolution is adjusted separately for each decay channel to reproduce the peak width in data, with corrections between 2.5\% and 4.7\%. The $M^2_{\mathrm{miss}}$ and $E_{\mathrm{ECL}}$ resolutions are studied using $\bar{B}\to D^{(*)}\ell^-\bar\nu_\ell$ candidates selected as the signal, but with the inverted requirement $q^2 < 3.5$~GeV$^2/c^2$ ($q^2$ sideband). 
 In this region, we find that simulation underestimates the width of the peak at $M^2_{\mathrm{miss}}=0$ by 3.9--14.4\%, depending on the $D^{(*)}$ mode. Corresponding corrections are applied with uncertainties dominated by the sideband sample size. No correction is required for the $E_{\mathrm{ECL}}$ resolution as the shape of the distribution matches the data in the $q^2$ sideband~\cite{SupplM}.

The candidates are divided into five exclusive samples, according to the charm meson originating from the $B_{\rm sig}$ candidate: $D^{*+}$, $D^{*0}_{[D^0\pi^0]}$, $D^{*0}_{[D^0\gamma]}$, $D^0$, and $D^{+}$. The $D^{*+}$ sample contains candidates with both $D^+\pi^0$ and $D^0\pi^+$.




In all samples, the expected composition is dominated by $\bar{B}\rightarrow D^{*} \ell^{-}\bar{\nu}_{\ell}$, contributing 70--85\% of the $D^{*}$ samples and 40--60\% of the $D$ samples through feed-down. The $\bar{B}\rightarrow D \ell^{-}\bar{\nu}_{\ell}$ contribution is expected to be the second-largest component in the $D^0$ (20\%) and $D^+$ (35\%) samples, and to contribute 1\% or less in the $D^{*}$ samples. Assuming LFU, $\bar{B}\rightarrow D^{*} \tau^{-}\bar{\nu}_{\tau}$ and $\bar{B}\rightarrow D \tau^{-}\bar{\nu}_{\tau}$ decays are expected to contribute about 4\% and 2\%, respectively, with the latter negligible in the $D^{*}$ samples except for $D^{*0}_{[D^0\gamma]}$ (1\%). Semileptonic backgrounds from $\bar{B}\rightarrow D^{**} \ell^{-}\bar{\nu}_{\ell}$ and gap modes contribute 7--15\%, depending on the sample. Hadronic $B$ decays contribute 3--7\%, arising primarily from $\bar{B} \rightarrow D^{(*)}\bar{D}_{s}^{(*)}$, $\bar{B} \rightarrow D^{(*)}\bar{D}^{(*)}K$, and $\bar{B} \rightarrow D^{(*)}n\pi(\pi^{0})$ decays, where $n\pi(\pi^{0})$ denotes any number of pions. Misreconstructed $D^{(*)}$ mesons contribute below 2\%, continuum is about 1\% or negligible except in the $D^+$ sample (2\%), and all other backgrounds total less than 0.1\%.

The main background sources, $\bar{B}\rightarrow D^{**} \ell^{-}\bar{\nu}_{\ell}$, gap modes, and hadronic $B$ decays, are only loosely constrained by existing measurements. Their modeling is validated with a dedicated control sample by reconstructing $\bar{B}\rightarrow D^* \pi^{0} \ell^{-}\bar{\nu}_{\ell}$ candidates. 
In this control sample, the $M^2_{\mathrm{miss}}$ and $E_{\mathrm{ECL}}$ distributions indicate a smaller gap-mode contribution in data than in the nominal model~\cite{SupplM}. 
The gap-mode yield is therefore left free in the signal fit, while alternative fits with a fixed yield of gap modes, up to twice that of the nominal model, are used to assign a systematic uncertainty. Hadronic $B$ decays are broadly consistent between data and simulation, except in the $D^{*0}_{[D^0\pi^0]}$ sample, where the data show lower yields~\cite{SupplM}; these differences are covered by systematic uncertainties. The $\bar{B}\rightarrow D^{**} \ell^{-}\bar{\nu}_{\ell}$ component is consistent between control data and simulation within the limited precision of existing measurements, which are propagated in the signal fit as a systematic uncertainty.

Misreconstructed $D^{(*)}$ decays that do not peak in the signal window, though a smaller contribution, are also validated. We use candidates that satisfy the selection but lie outside the $D$-mass or $\Delta M$ signal windows ($D^{(*)}$ sidebands). The data yield is consistent with simulation, and we assign no systematic uncertainties. In addition, a fraction of 8--13\% of misreconstructed $D^*$ candidates peaks in $\Delta M$ across the $D^*$ fit components. Since the corresponding $E_{\mathrm{ECL}}$ and $M_{\mathrm{miss}}^{2}$ distributions differ only slightly from those of correctly reconstructed candidates, we merge them within each component. The fractions are validated in the $q^2$ sideband by fitting the $E_{\mathrm{ECL}}$ distribution with free correctly reconstructed  and misreconstructed components. The statistical uncertainty on the fractions estimated in the $D^{(*)}$ sidebands is propagated as a systematic uncertainty. 
The values of $R(D^{(*)})$ are determined from an extended maximum-likelihood fit to the two-dimensional distribution of $E_{\mathrm{ECL}}$ and $M_{\mathrm{miss}}^{2}$, with $E_{\mathrm{ECL}} \in [0,2]$ GeV divided into 10 bins and $M_{\mathrm{miss}}^{2} \in [-2,10]$ GeV$^2/c^4$ divided into 14 bins. The fit is performed simultaneously across the five $D^{(*)}$ samples.

In each sample, the fit components are: the signal $\bar{B}\rightarrow D^{(*)} \tau^{-}\bar{\nu}_{\tau}$ and normalization $\bar{B}\rightarrow D^{(*)} \ell^{-}\bar{\nu}_{\ell}$ processes, backgrounds from $\bar{B}\rightarrow D^{**} \tau^{-}\bar{\nu}_{\tau}$ and $\bar{B}\rightarrow D^{**} \ell^{-}\bar{\nu}_{\ell}$, gap modes for $\tau$ leptons and light leptons, hadronic $B$ decays, continuum, and other minor backgrounds not included in the above categories and considered together as a single component. Each component is modeled by a probability density function (PDF), derived from simulated histogram templates including control-data corrections, with an associated yield.

The signal yield is expressed as 
$N_{D^{(*)}\tau\nu} =R(D^{(*)})N_{D^{(*)}\ell\nu}(\varepsilon_{D^{(*)}\tau\nu}/\varepsilon_{D^{(*)}\ell\nu})/2$, where the factor of two accounts for averaging over the two light-lepton families, and $\varepsilon_{D^{(*)}\tau\nu}$ and $\varepsilon_{D^{(*)}\ell\nu}$ are the signal and normalization selection efficiencies. Both efficiencies include the branching fractions of the reconstructed $D^{(*)}$ decay chains, and $\varepsilon_{D^{(*)}\tau\nu}$ additionally includes $\mathcal B(\tau^-\to\ell^-\bar\nu_\ell\nu_\tau)$.  
To enhance sensitivity, we assume isospin symmetry for charged and neutral $B$ meson decays, and treat $R(D^{(*)})= R(D^{(*)0})=R(D^{(*)+})$ as a single free parameter common to all samples. 

The normalization yields are given by 
 $N_{D^{(*)} \ell \nu} = 4\mathcal{B} (\bar{B} \rightarrow D^{(*)} \ell^{-} \bar{\nu}_{\ell})  N_{\Upsilon(\mathrm{4S})} f_i\, \varepsilon_{D^{(*)} \ell \nu}$,
where the branching fraction $\mathcal{B}(\bar{B} \rightarrow D^{(*)}\ell^{-} \bar{\nu}_{\ell})$ is a free parameter, while the number of $\Upsilon(4S)$ events, $N_{\Upsilon(4S)}=(387.1\pm5.6)\times 10^6$, and the branching fractions $f_i$ for $\Upsilon(4S)$  decays into neutral or charged $B$ meson pairs, $f_{00}=0.4861\pm0.0080$ and $f_{+-}=0.5113\pm0.0108$~\cite{HFLAV:2024}, are Gaussian-constrained parameters. 
The overall factor of four accounts for the two $B$ mesons produced per $\Upsilon(4S)$ decay and for averaging over the light-lepton families.

 The signal efficiencies and background yields have specific values in each $D^{(*)}$ sample. The efficiencies, as well as the yields of continuum and other backgrounds, are fixed from simulation and include the control-data corrections, while all other yields float in the fit. The procedure is validated with ensembles of simulated experiments.

\begin{figure*}
    \centering
        \begin{minipage}{1.0\linewidth}
            \includegraphics[width=\linewidth]{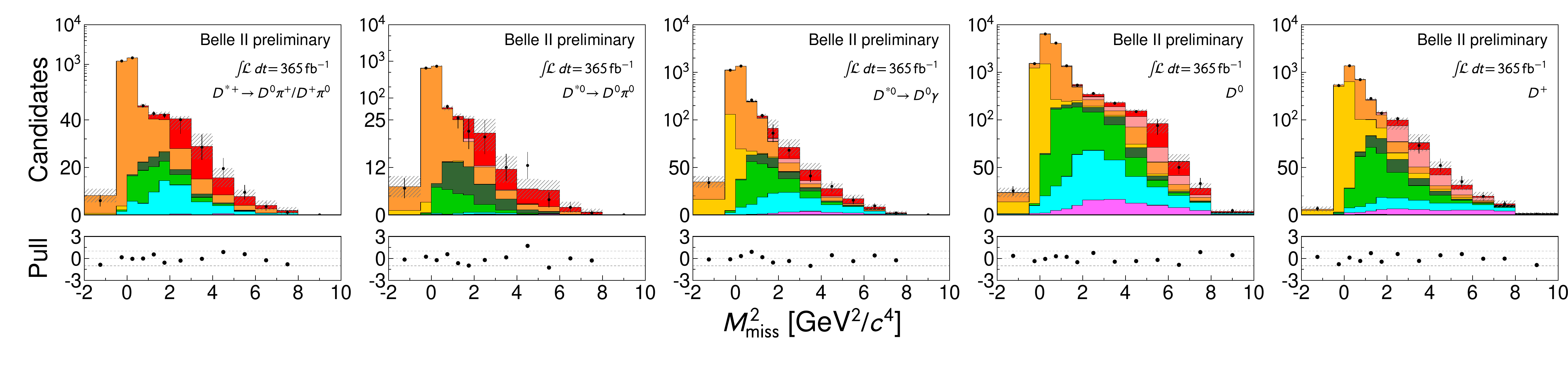}
        \end{minipage} \\
        \begin{minipage}{1.0\linewidth}
            \includegraphics[width=\linewidth]{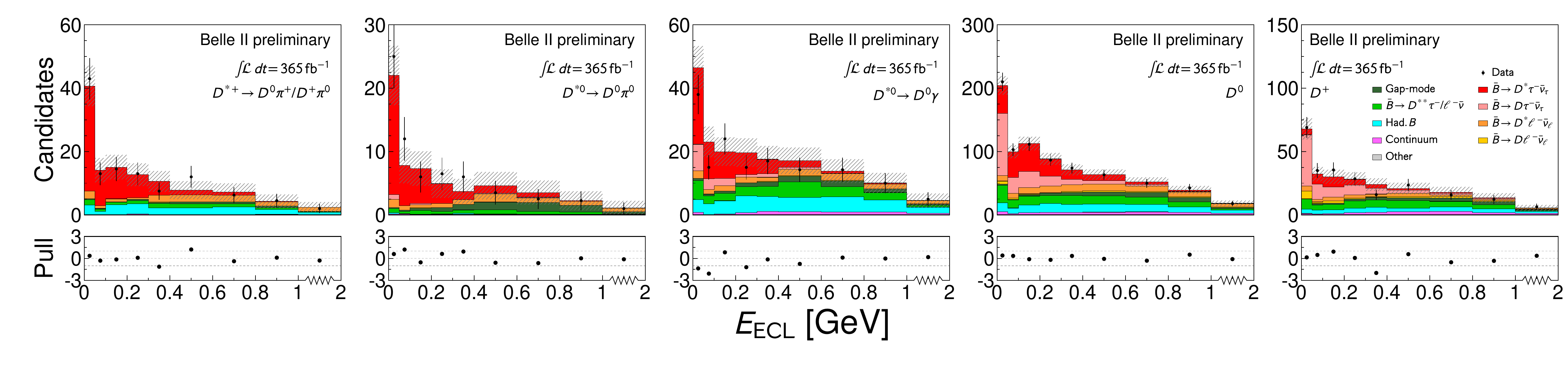}
        \end{minipage}
    \caption{Distributions of (top row)  $M^2_{\mathrm{miss}}$ and (bottom row) $E_{\mathrm{ECL}}$ with the fit results overlaid for the five samples (from left to right) $D^{*+}$, $D^{*0}_{[D^0\pi^0]}$, $D^{*0}_{[D^0\gamma]}$, $D^0$, and $D^{+}$. The hatched band represents the statistical uncertainty in the simulated prediction. To enhance the signal contribution in the $E_{\mathrm{ECL}}$ distributions, the requirement $M^2_{\mathrm{miss}} > 1.5~\mathrm{GeV}^2/c^4$ is applied for the $D^*$ modes and $M^2_{\mathrm{miss}} > 2~\mathrm{GeV}^2/c^4$ for the $D$ modes. The lower panel of each distribution shows the difference between the data and the fit normalized to the data uncertainty (pull).
    }
    \label{fig:fit_result}
\end{figure*}

The branching fractions of $\bar{B}\rightarrow D^{*} \ell^{-}\bar{\nu}_{\ell}$ and $\bar{B}\rightarrow D \ell^{-}\bar{\nu}_{\ell}$ are consistent with world averages~\cite{HFLAV:2024} within 1.2 and 0.1 standard deviations, respectively, when accounting for statistical and systematic uncertainties. The yields of $\bar{B}\rightarrow D^{**} \tau/\ell^{-}\bar{\nu}$, gap modes, and hadronic $B$ decays correspond to branching fractions consistent, within statistical and systematic uncertainties, with those obtained in the $\bar{B}\rightarrow D^* \pi^{0} \ell^{-}\bar{\nu}_{\ell}$ control sample~\cite{SupplM}. Figure~\ref{fig:fit_result} shows the $M^2_{\mathrm{miss}}$ and $E_{\mathrm{ECL}}$ distributions with the fit results overlaid.

\begin{figure}
    \centering
        \begin{minipage}{0.9\linewidth}
            \includegraphics[width=\linewidth]{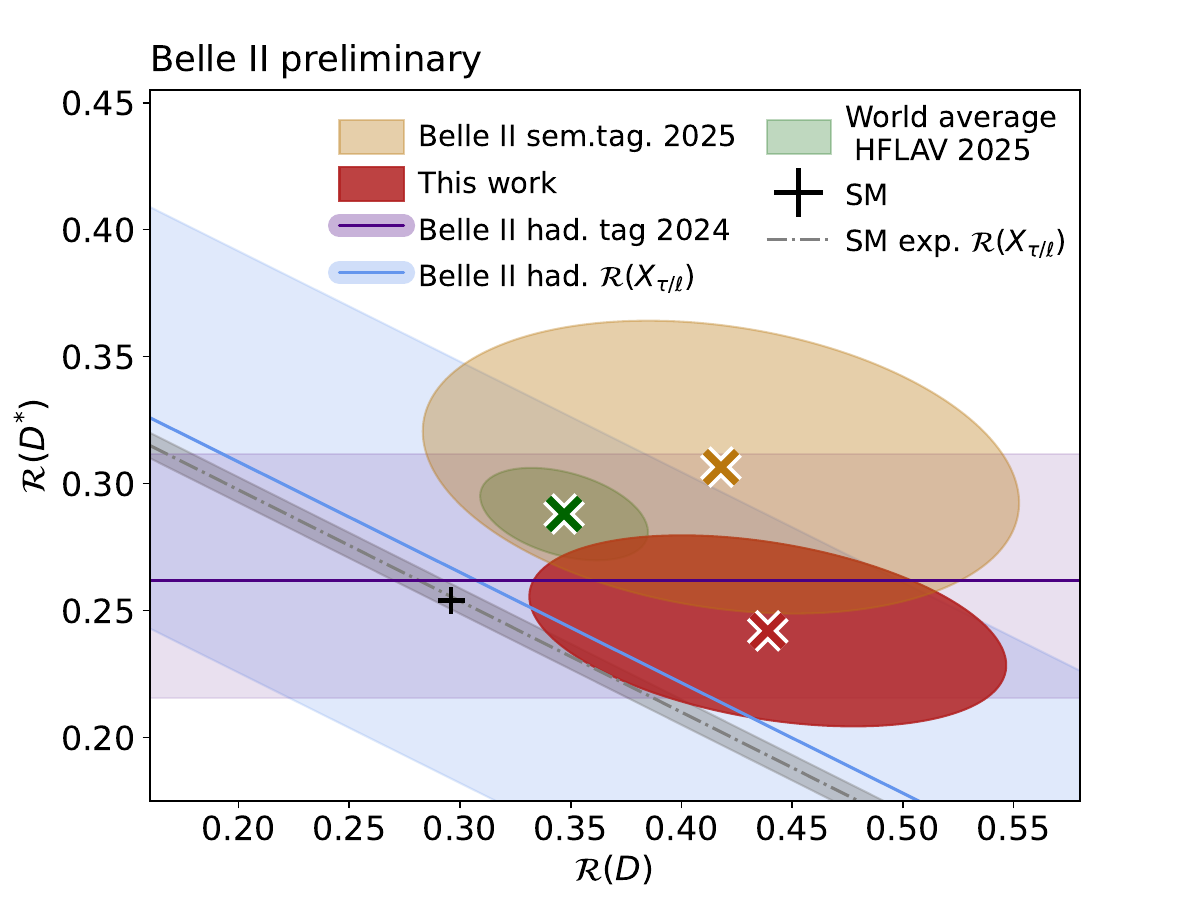}
        \end{minipage} 
    \caption{The measured $R(D)$ and $R(D^{*})$ values (red marker) and their 68.3\% confidence region (red ellipse) are shown together with the SM prediction (black marker), the world average~\cite{HFLAV:2024} (green ellipse), the superseded Belle II measurement of $R(D^{*})$ with hadronic tagging~\cite{Belle2RDstar} (dark violet band), the Belle II measurement of $R(D^{(*)})$ with semileptonic $B$-tagging~\cite{Belle2RDRDstar} (yellow ellipse), and the constraint from the Belle II measurement of $R(X_{\tau/\ell})$~\cite{Belle2RX} (blue band). The grey band indicates the SM expectation for $R(X_{\tau/\ell})$ from Ref.~\cite{Belle2RX}.
    }
    \label{fig:2D}
\end{figure}

\begin{table}
    \centering
    \caption{
       Summary of the relative uncertainties on $R(D^{(*)})$ and their correlation coefficients ($\rho$). The description of each systematic source is provided in the text. 
    }
    \resizebox{\columnwidth}{!}{
    \begin{tabular}{lccc}
        \toprule
        \toprule
         Source & $R(D^{*})$ & $R(D)$ & $\rho$ \\
        \midrule
        Simulation sample size                              & 4.8\% & 8.4\% & $-0.44$ \\
        Gap-mode branching fractions
           & 2.7\% & 3.1\% & $\phantom{-}0.00$  \\ 
        {$\bar{B} \rightarrow D^{**}\tau^-/(\ell^{-})\bar{\nu}$ branching fractions}   
                                                     & 0.3\% & 1.3\% & $\phantom{-}0.25$  \\
        Hadronic $B$ decay branching fractions       & 1.6\% & 1.5\% & $-0.26$  \\
        Form factors                                 & 0.5\% & 0.9\% & $-0.70$  \\
        Misreconstructed $D^{(*)}$ candidates      & 0.5\% & 1.2\% &  $\phantom{-}0.00$  \\
        Continuum background                         & 2.4\% & 2.1\% & $\phantom{-}0.93$   \\
        Fit biases                                   & 0.3\% & 1.2\% &  $\phantom{-}0.00$  \\
        Low-momentum $\pi^0$ and $\gamma$ efficiencies      & 2.2\% & 2.4\% & $\phantom{-}0.99$  \\
        $B$-tagging efficiency                        & 0.9\% & 2.5\% & $-1.00$   \\
        Other efficiency corrections                 & 0.7\% & 1.4\% & $\phantom{-}0.92$   \\
        {$M^{2}_{\mathrm{miss}}$ resolution}         & 0.5\% & 0.8\% &  $\phantom{-}0.48$ \\
        \midrule 
        Total systematic uncertainty                 & 6.7\% & 10.3\% & $-0.20$ \\ 
        \midrule 
        Statistical uncertainty                & 7.9\% & 12.5\% & $-0.40$ \\
        \bottomrule
        \bottomrule
    \end{tabular}
    }
    \label{tab:sys_summary}
\end{table}

Systematic uncertainties are evaluated to account for the limited size of the simulated samples, background validation, branching-fraction and modeling assumptions, fit biases, efficiency corrections and resolution mismodeling. All uncertainties on $R(D^{*})$ and $R(D)$, together with their correlations, are reported in Table~\ref{tab:sys_summary}.  

In general, we refit the data under several variations of the PDFs associated with a given systematic source,  and take the standard deviation of the resulting $R(D^{(*)})$ distribution as the systematic uncertainty. When this distribution is non-Gaussian, we assign as the uncertainty the larger of the positive- and negative-side widths.  These asymmetric distributions are  observed for the systematic uncertainties associated with: gap-mode and $\bar{B} \rightarrow D^{**} \tau^{-} \bar{\nu}_{\tau}$ branching fractions, continuum background, and fit biases. 
The total systematic uncertainty is obtained by simultaneously varying all sources.

The limited size of the simulated samples is addressed using bootstrap resampling~\cite{BS:1988} to vary both the PDFs and the signal efficiencies. For each component and $D^{(*)}$ sample, we fluctuate the event counts according to Poisson distributions and resample the simulated events with replacement to derive alternative PDFs and efficiencies.  

The yield of gap modes is floated in the fit and is found to be 42--82\% smaller in data than in simulation, depending on the $D^{*}$ samples. This result is consistent with the $\bar{B} \rightarrow D^{*} \pi^0 \ell^{-} \bar{\nu}_{\ell}$ control sample. To cover this poorly known background, we perform alternative fits either by fixing the yield to values corresponding to 0\% and 200\% of the estimated branching fraction or by varying it uniformly within this range. 


For $\bar{B} \rightarrow D^{**} \ell^{-} \bar{\nu}_{\ell}$ decays, we vary the branching fractions according to Gaussian distributions using their measured values and uncertainties. For $\bar{B} \rightarrow D^{**} \tau^{-} \bar{\nu}_{\tau}$ decays, each branching fraction is varied uniformly between 0 and 200\% of the estimated value. The composition of hadronic $B$ decays is treated analogously, with Gaussian variations for measured modes and uniform variations otherwise. Semileptonic form-factor parameters are varied according to Gaussian distributions, including correlations where available. 
We propagate to the PDFs the uncertainty on the fraction of misreconstructed $D^*$ mesons peaking in $\Delta M$, estimated in the $q^2$ sideband.


The continuum yield is fixed in the fit, and its uncertainty is evaluated by varying it uniformly between zero and twice the nominal value. 



Fit biases are studied using ensembles of simulated data sets fitted with the nominal model. 

Experimental uncertainties arise from efficiency corrections for tracking, particle identification, photon and $\pi^0$ reconstruction, and $B$-tagging efficiency, derived from control samples. 
 Efficiency-correction uncertainties are generally small; the largest arise from $B$-tagging and low-momentum $\pi^0$ and $\gamma$ reconstruction, which are reported separately from the other efficiency corrections in Table~\ref{tab:sys_summary}. 
All corrections are assumed to be correlated between $\bar{B} \rightarrow D^{(*)} \ell^{-} \bar{\nu}_{\ell}$ and $\bar{B} \rightarrow D^{(*)} \tau^{-} \bar{\nu}_{\tau}$ decays.  The $B_\text{tag}$ efficiency is further validated in $q^2$ sidebands. In these regions, data and simulation show consistent lepton-momentum spectra; however, we find that efficiencies for the $\bar{B} \rightarrow D \pi\pi\pi\pi^0$ tagging mode differ between $\bar{B} \rightarrow D^{(*)} \tau^{-} \bar{\nu}_{\tau}$ and $\bar{B} \rightarrow D^{(*)} \ell^{-} \bar{\nu}_{\ell}$ decays. These differences are propagated as a systematic uncertainty. 

Additional systematic uncertainties are assigned from the calibration of the $M^2_{\mathrm{miss}}$ resolution, propagated to both the PDF shapes and the efficiencies. 

Uncertainties on $N_{\Upsilon(\mathrm{4S})}$, $f_{00}$, and $f_{+-}$ cancel in the ratio and are therefore neglected.

To assess the stability of the result, we compare the $R(D^{(*)})$ values obtained in various independent subsamples. We split the data according to $D^0$ and $D^+$ mesons, electron and muon channels, different run periods, and whether the $q^2$ sideband is included to simultaneously fit the branching fractions of $\bar{B} \rightarrow D^{(*)} \ell^{-} \bar{\nu}_{\ell}$ and $R(D^{(*)})$~\cite{SupplM}. The resulting $R(D^{(*)})$ values are consistent with statistical fluctuations and no additional systematic uncertainty is assigned from these stability tests.

After incorporating all systematic uncertainties, we obtain  
\begin{align}
    R(D^{*}) &= {0.242 \pm 0.019 \mathrm{(stat)} \pm 0.016 \mathrm{(syst)}}\,, \\ 
    R(D)     &= {0.439 \pm 0.055 \mathrm{(stat)} \pm 0.046 \mathrm{(syst)}}\,. 
\end{align}
The correlation coefficients of the uncertainties are $-0.40$(stat) and $-0.20$(syst). 

The results are consistent with the SM predictions within $0.5\sigma$ for $R(D^{*})$ and $2.0\sigma$ for $R(D)$, where the SM reference is taken as the arithmetic average reported in Ref.~\cite{HFLAV:2024}. The combined measurement of $R(D)$ and $R(D^{*})$ agrees with the SM within $1.5\sigma$ and with the current world average~\cite{HFLAV:2024} within $1.3\sigma$. Figure~\ref{fig:2D} shows the 68.3\% confidence-level region in the $R(D)$--$R(D^{*})$ plane from this analysis, compared with previous Belle~II measurements~\cite{Belle2RDstar, Belle2RDRDstar, Belle2RX}, the world average, and the SM predictions. This measurement of $R(D^{*})$ supersedes our earlier result~\cite{Belle2RDstar}, improving the precision by a factor of two. 
We obtain the most precise determination of $R(D^{(*)})$ with hadronic tagging, reaching a precision comparable to that of the best simultaneous measurements of these ratios~\cite{BelleDtaunu4, LHCb2023}.  



\begin{acknowledgments}
\input{acknowledgements-b2}
\end{acknowledgments}

\bibliography{references}  

\end{document}


\title{Test of lepton flavor universality with \boldmath{$\bar{B} \rightarrow D^{(*)} \tau^{-} \bar{\nu}_{\tau}$} \\ and \boldmath{$\bar{B} \rightarrow D^{(*)} \ell^{-} \bar{\nu}_{\ell}$} decays at Belle~II\\
\vspace{0.5cm}
(The Belle II Collaboration)}

\maketitle

\section{Supplemental Material}

\subsection{Validation with the \boldmath{$q^2$} sideband}

\begin{figure*}[h]
    \centering 
        \begin{minipage}{1.0\linewidth}
            \includegraphics[width=\linewidth]{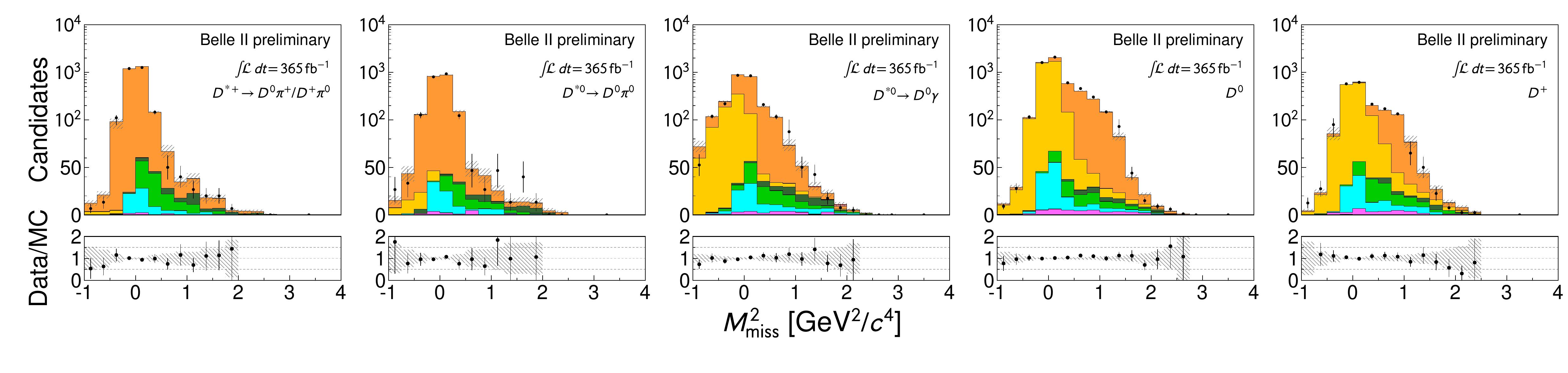}
        \end{minipage} \\
        \begin{minipage}{1.0\linewidth}
            \includegraphics[width=\linewidth]{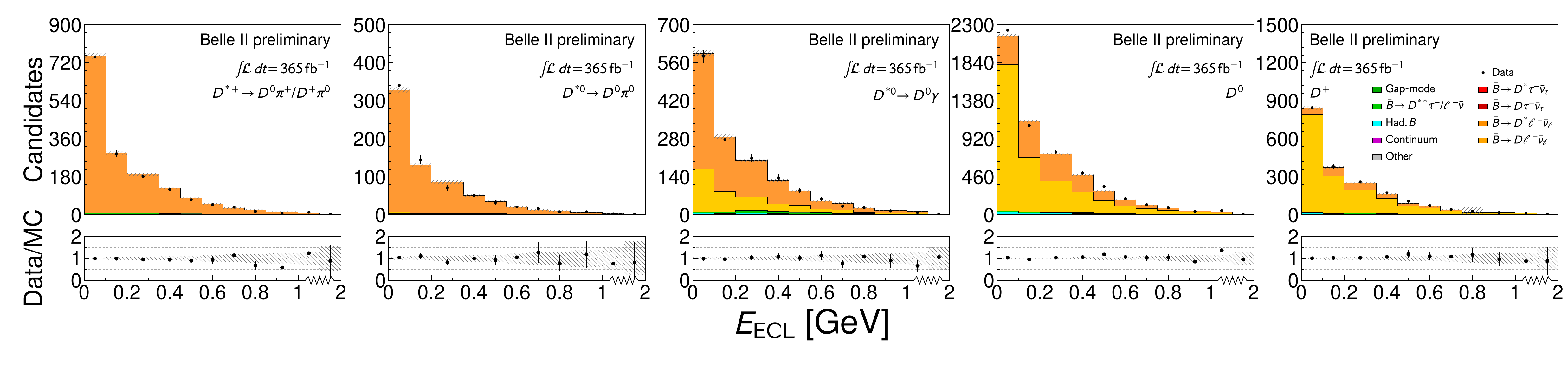}
        \end{minipage}
    \caption{Distributions of (top row) $M^2_{\mathrm{miss}}$ and (bottom row) $E_{\mathrm{ECL}}$ for the five samples (from left to right) $D^{*+}$, $D^{*0}_{[D^0\pi^0]}$, $D^{*0}_{[D^0\gamma]}$, $D^0$, and $D^{+}$ in the $q^2<$ 3.5~GeV$^2/c^2$ sideband. The points represent the observed data, while the stacked histograms represent the simulated contributions from the different processes. The lower panel of each distribution shows the ratio of the observed data to the simulation prediction (Data/MC). The hatched band represents the statistical uncertainty in the simulated prediction.
    }
    \label{fig:q2-prefit}
\end{figure*}

To validate simulation modeling, we compare the yields and distributions of the decays $\bar{B}\rightarrow D^{(*)} \ell^{-}\bar{\nu}_{\ell}$ predicted by simulation with those observed in the data in the $q^{2}$ sideband, defined by $q^2<3.5\,\mathrm{GeV}^2/c^2$.  
Figure~\ref{fig:q2-prefit} shows the $M^2_{\mathrm{miss}}$ and $E_{\mathrm{ECL}}$ distributions. 
The predicted yields are consistent with the data when statistical and systematic uncertainties are considered. 
Considering statistical uncertainties only, the largest deviation is approximately $2\sigma$ and occurs in the $D^{+}\to D^+\pi^0$ sample. 
The $E_{\mathrm{ECL}}$ resolution in simulation is consistent with that observed in data, while the $M^2_{\mathrm{miss}}$ resolution agrees with data after applying the correction described in the main text. 

\begin{figure*}[h]
    \centering 
        \begin{minipage}{1.0\linewidth}
            \includegraphics[width=\linewidth]{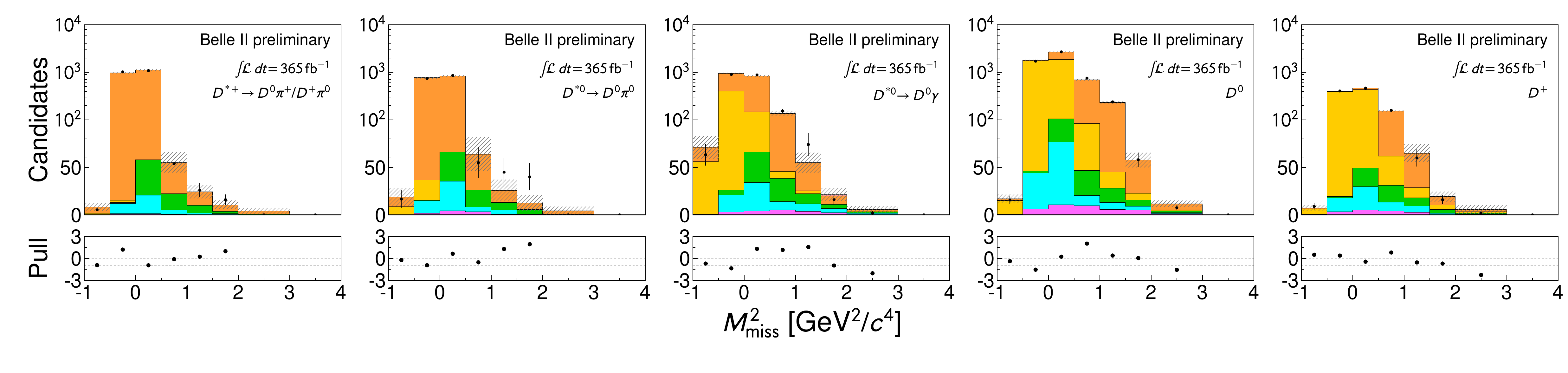}
        \end{minipage} \\
        \begin{minipage}{1.0\linewidth}
            \includegraphics[width=\linewidth]{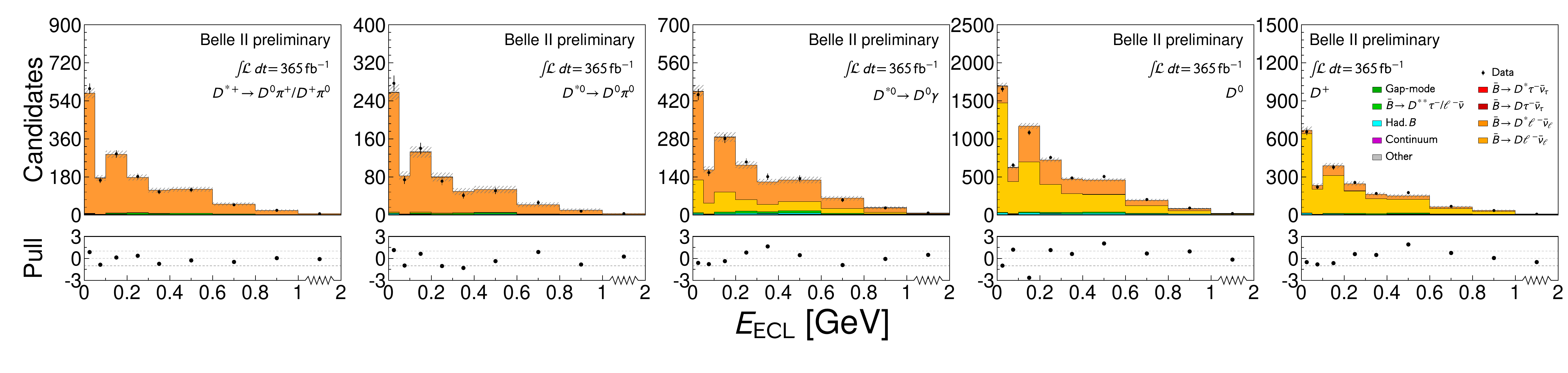}
        \end{minipage}
    \caption{Distributions of (top row) $M^2_{\mathrm{miss}}$ and (bottom row) $E_{\mathrm{ECL}}$ with fit results overlaid for the five samples (from left to right) $D^{*+}$, $D^{*0}_{[D^0\pi^0]}$, $D^{*0}_{[D^0\gamma]}$, $D^0$, and $D^{+}$ in the $q^2<$ 3.5~GeV$^2/c^2$ sideband. The points represent the observed data, while the stacked histograms represent the fitted component yields. The hatched band represents the statistical uncertainty in the simulation. The lower panel of each distribution shows the difference between the data and the fitted expectation divided by the data uncertainty (pull).
    }
    \label{fig:q2-postfit}
\end{figure*}

As an additional consistency check, we perform a fit restricted to the $q^2$ sideband. Because this region contains only a small signal contribution and therefore has little sensitivity to $R(D^{(*)})$, these ratios are fixed to their standard model values~\cite{HFLAV:2024}. The background parameters that float in the nominal signal fit are fixed to their expected values. 
The branching fractions of $\bar B\to D^{(*)}\ell^-\bar\nu_\ell$ are allowed to vary, and their values from the fit are found to be consistent with world averages~\cite{HFLAV:2024} within the statistical and systematic uncertainties. 
Figure~\ref{fig:q2-postfit} presents the distributions with the fit results overlaid. 

We also perform a combined fit to the signal region and the $q^2$ sideband, simultaneously determining $R(D^{(*)})$ and the $\bar B\to D^{(*)}\ell^-\bar\nu_\ell$ branching fractions. We obtain $R(D^{*}) = 0.246 \pm 0.020 \mathrm{(stat)}$ and $R(D)= 0.430 \pm 0.052 \mathrm{(stat)}$, which differ from the nominal results by $1.7\%$ and $2.1\%$, respectively. The normalization branching fractions obtained from the combined fit remain consistent with the world averages within their uncertainties. No additional systematic uncertainty is assigned on the basis of this stability test.

\subsection{Validation with the \boldmath{$\bar{B}\rightarrow D^* \pi^{0} \ell^{-}\bar{\nu}_{\ell}$} control sample}
According to the simulation, $\bar{B}\rightarrow D^{**} \ell^{-}\bar{\nu}_{\ell}$ and gap modes contribute between 7--15\% of the selected events depending on the samples, while hadronic $B$ decays contribute 3--7\%. 
We validate the modeling of these backgrounds using a control sample obtained by reconstructing $\bar{B}\rightarrow D^* \pi^{0} \ell^{-}\bar{\nu}_{\ell}$ candidates. 

\begin{figure*}[h]
    \centering
        \begin{minipage}{1.0\linewidth}
            \includegraphics[width=\linewidth]{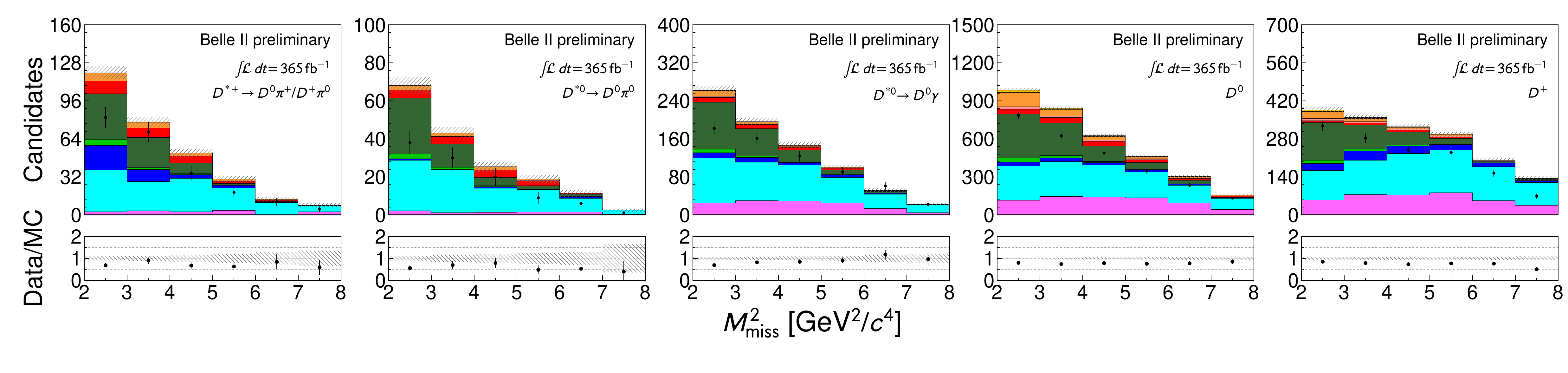}
        \end{minipage} \\
        \begin{minipage}{1.0\linewidth}
            \includegraphics[width=\linewidth]{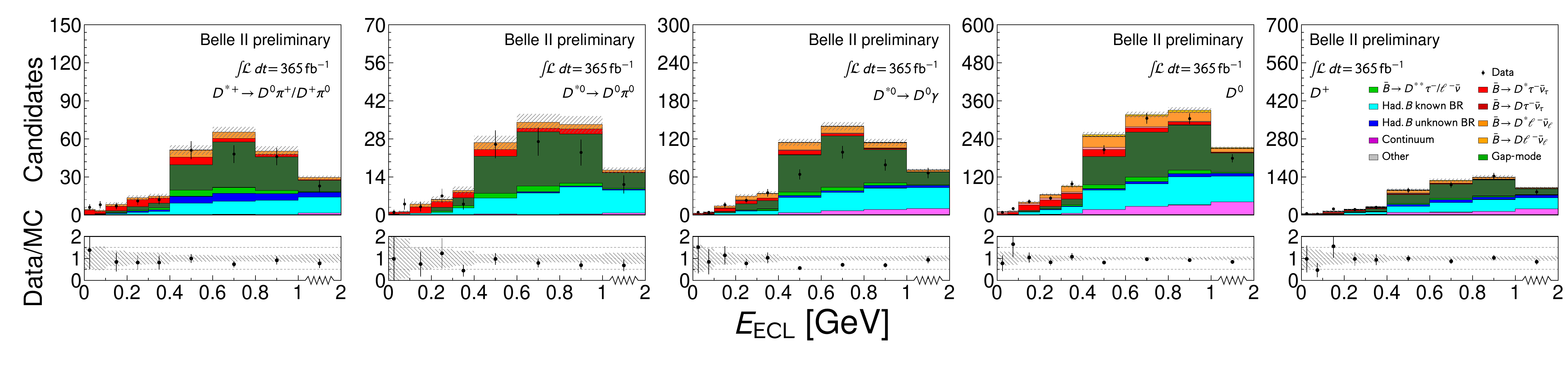}
        \end{minipage}
    \caption{Distributions of (top row) $M^2_{\mathrm{miss}}$ and (bottom row) $E_{\mathrm{ECL}}$ for the five samples (from left to right) $D^{*+}$, $D^{*0}_{[D^0\pi^0]}$, $D^{*0}_{[D^0\gamma]}$, $D^0$, and $D^{+}$ in the $\bar{B} \rightarrow D^{*} \pi^0 \ell^{-} \bar{\nu}_{\ell}$ control sample. To enhance the $\bar{B} \rightarrow D^{**} \ell^{-} \bar{\nu}_{\ell}$ contribution in the $E_{\mathrm{ECL}}$ distributions, the requirements $1 < M^2_{\mathrm{miss}} <5~\mathrm{GeV}^2/c^{4}$ and $2 < M_{\mathrm{miss}}^{2} < 5~\mathrm{GeV}^2/c^{4}$ are applied to the $D^*$ modes and $D$ modes, respectively. For the $M^2_{\mathrm{miss}}$ distributions, events satisfying $M^2_{\mathrm{miss}} > 2~\mathrm{GeV}^2/c^4$ are shown for the $D^{(*)}$ modes. The points represent the observed data, while the stacked histograms represent the simulated contributions from the different processes. The lower panel of each distribution shows the ratio of the observed data to the simulation prediction (Data/MC). The hatched band represents the statistical uncertainty in the simulated prediction.}
    \label{fig:pi0vetoPrefit}
\end{figure*}

As illustrated in Fig.~\ref{fig:pi0vetoPrefit}, we inspect the $E_{\mathrm{ECL}}$ and $M_{\rm miss}^2$ distributions. For the $E_{\rm ECL}$ distributions, we require $1<M_{\rm miss}^2<5\,\mathrm{GeV}^2/c^4$ in the $D^*$ samples and $2<M_{\rm miss}^2<5\,\mathrm{GeV}^2/c^4$ in the $D$ samples. For the $M_{\rm miss}^2$ distributions, we require $M_{\rm miss}^2>2\,\mathrm{GeV}^2/c^4$ in all $D^{(*)}$ samples. In these selected subsamples, gap modes and hadronic $B$ decays constitute the dominant contributions. Depending on the $D^{(*)}$ sample, the number of selected events in data is 10--40\% lower than predicted by simulation.

\begin{figure*}[h]
    \centering
        \begin{minipage}{1.0\linewidth}
            \includegraphics[width=\linewidth]{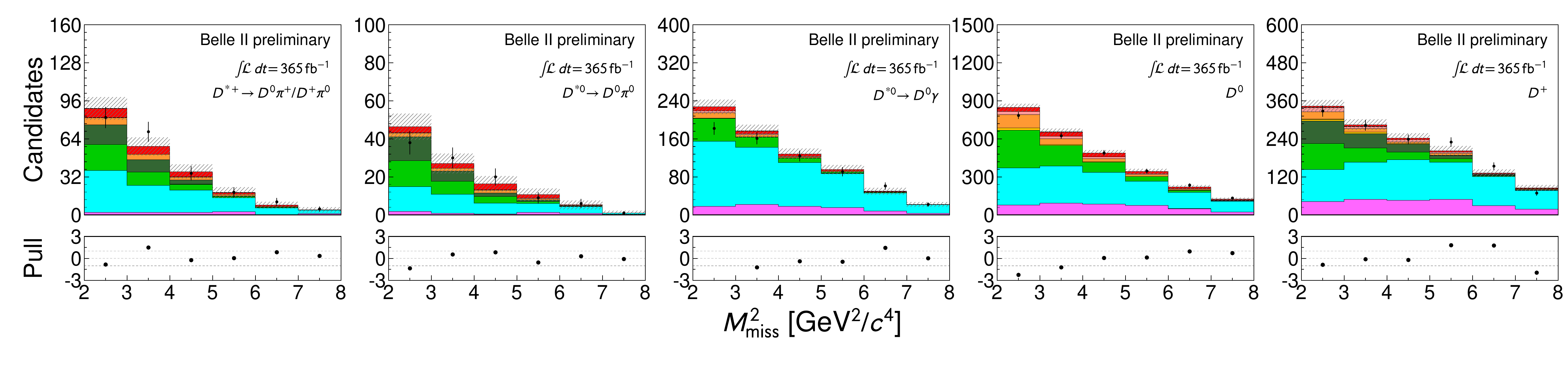}
        \end{minipage} \\
        \begin{minipage}{1.0\linewidth}
            \includegraphics[width=\linewidth]{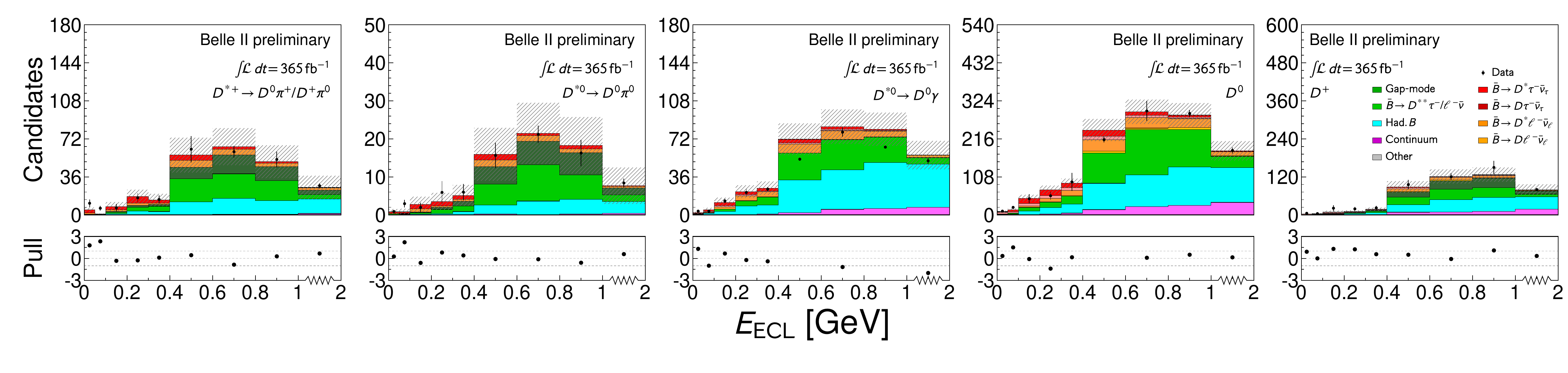}
        \end{minipage}
    \caption{Distributions of (top row) $M^2_{\mathrm{miss}}$ and (bottom row) $E_{\mathrm{ECL}}$ with fit results overlaid for the five samples (from left to right) $D^{*+}$, $D^{*0}_{[D^0\pi^0]}$, $D^{*0}_{[D^0\gamma]}$, $D^0$, and $D^{+}$ in the $\bar{B} \rightarrow D^{*} \pi^0 \ell^{-} \bar{\nu}_{\ell}$ control sample. Hadronic $B$ decays with measured and unmeasured branching fractions are summed together and shown as a single component (Hadronic $B$) in the fit projection. To enhance the $\bar{B} \rightarrow D^{**} \ell^{-} \bar{\nu}_{\ell}$ contribution in the $E_{\mathrm{ECL}}$ distributions, the requirements of $1 < M^2_{\mathrm{miss}} <5~\mathrm{GeV}^2/c^{4}$ and $2 < M_{\mathrm{miss}}^{2} < 5~\mathrm{GeV}^2/c^{4}$ are applied to the $D^*$ modes and $D$ modes, respectively. For the $M^2_{\mathrm{miss}}$ distributions, events satisfying $M^2_{\mathrm{miss}} > 2~\mathrm{GeV}^2/c^4$ are shown for the $D^{(*)}$ modes. The points represent the observed data, while the stacked histograms represent the fitted component yields. The hatched band represents the statistical uncertainty in the simulation. The lower panel of each distribution shows the difference between the data and the fitted expectation divided by the data uncertainty (pull).}
    \label{fig:pi0vetoPostfit}
\end{figure*}

To constrain the normalizations of $\bar{B}\rightarrow D^{**} \ell^{-}\bar{\nu}_{\ell}$ decays, gap modes, and hadronic $B$ decays, we perform a simultaneous fit across the five $D^{(*)}$ samples to the two-dimensional $E_{\rm ECL}$ and $M_{\rm miss}^2$ distributions. The $\bar{B}\rightarrow D^{**} \ell^{-}\bar{\nu}_{\ell}$ and gap-mode contributions are treated as independent components. Hadronic $B$ decays are separated into modes with measured and unmeasured branching fractions because different constraints and systematic variations are assigned to the two categories. Figure~\ref{fig:pi0vetoPostfit} shows the resulting fit projections.

We compare the fitted background normalizations with the simulation predictions in the signal region and in the $\bar{B}\rightarrow D^* \pi^{0} \ell^{-}\bar{\nu}_{\ell}$ control sample. In the signal region, the fitted gap-mode yield is 42--82\% lower than the simulation prediction, depending on the $D^{(*)}$ sample. The corresponding normalization factors are consistent with those obtained in the control sample. The fitted hadronic-$B$-decay normalizations are broadly consistent with the simulation prediction in both fits, except in the $D^{*0}_{[D^0\pi^0]}$ sample, where the normalization is $3.1\sigma$ below the prediction in the signal region and $0.8\sigma$ below it in the control sample.


The systematic uncertainties associated with the gap-modes and hadronic $B$ decays explained in the main text cover the discrepancies observed between data and simulation, and no additional systematic uncertainty is assigned.



\subsection{Nominal fit yields}
The component yields obtained from the nominal fit used to determine $R(D^{(*)})$ are summarized in Table~\ref{tab:finalfitYields}.  

\onecolumngrid

\begin{table*}[h]
    \centering
    \caption{Component yields in each exclusive $D^{(*)}$ sample determined from the fit. Uncertainties are statistical when present; otherwise, the reported value is fixed in the fit.}
    \begin{tabularx}{0.8\textwidth}{l
                    S[table-format=7.0(2)]   
                    S[table-format=7.0(2)]   
                    S[table-format=7.0(2)]   
                    S[table-format=8.0(3)]   
                    S[table-format=7.0(2)]  
    }
        \toprule
        \toprule
        & 
        \multicolumn{1}{c}{${D^{*}}^{+}$} &
        \multicolumn{1}{c}{${D^{*0}_{[D^{0}\pi^{0}]}} $} &
        \multicolumn{1}{c}{${D^{*0}_{[ D^{0}\gamma]}} $} &
        \multicolumn{1}{c}{$D^{0}$} &
        \multicolumn{1}{c}{$D^{+}$} 
        \\
        \midrule
 

        $\bar{B}\rightarrow D^{*} \tau^{-}\bar{\nu}_{\tau}$              & 124.4  +- 9.8  & 64.2   +- 5.0  & 106.5  +- 8.4 & 380.1  +- 29.8   & 63.4   +- 5.0  \\
        $\bar{B}\rightarrow D \tau^{-}\bar{\nu}_{\tau}$                  & <0.1  & 2.1    +- 0.3  & 31.5   +- 3.9 & 340.0   +- 42.6   & 147.8  +- 18.5 \\
        $\bar{B}\rightarrow D^{*} \ell^{-}\bar{\nu}_{\ell}$              & 2898.6 +- 65.8 & 1417.2 +- 32.2 & 2598.0 +- 58.8 & 10638.1 +- 235.4 & 1624.3 +- 36.9 \\
        $\bar{B}\rightarrow D \ell^{-}\bar{\nu}_{\ell}$                  & 3.8  +- 0.1   & 6.8  +- 0.2  & 214.8 +- 6.5  & 2881.9 +- 86.9  & 1225.6 +- 37.0  \\
        $\bar{B}\rightarrow D^{**} \tau^{-}/\ell^{-}\bar{\nu}$           & 60.1 +- 31.2  & 28.9 +- 23.9 & 190.5 +- 58.3 & 731.1 +- 134.0 & 294.2 +- 63.5  \\
        Gap modes                                       & 21.4 +- 24.8  & 41.6 +- 15.1 & 53.4  +- 49.6 & 240.3 +- 93.5  & 89.5  +- 46.4  \\
        Hadronic $B$ decays                             & 86.0 +- 25.3  & 3.5 +- 12.2  & 165.6 +- 38.2 & 469.2 +- 94.6  & 125.6 +- 25.3  \\
        Continuum                                       & 2.9    & 2.2    & 33.0   & 181.0  & 84.0    \\
        Other                                           & 0.2    & 0.0    & 0.1    & 0.8      & 4.2   \\
        \midrule
        Total                                           & 3197.4 & 1566.5 & 3393.4 & 15862.5 & 3658.7  \\
        
        \bottomrule
        \bottomrule
    \end{tabularx}
    \label{tab:finalfitYields}
\end{table*}

\bibliography{supplreferences}

%% file: definitions.tex
\usepackage[dvipdfmx]{graphicx} 

\usepackage{wrapfig}
\usepackage[caption=false]{subfig}
\usepackage{epstopdf} 
\usepackage{xcolor} 
\usepackage{amssymb} 
\usepackage{hyperref} 
\usepackage[utf8]{inputenc} 
\usepackage[english]{babel} 
\usepackage{blindtext} 
\usepackage{subfiles}
\usepackage{multirow}
\usepackage{booktabs}
\usepackage{bm}
\usepackage{cases}
\usepackage {autobreak}
\usepackage{tabularx}
\usepackage{comment}
\usepackage{orcidlink} 

\graphicspath{{Figures/}} 

\makeatletter
\renewcommand{\p@subsection}{}
\renewcommand{\p@subsubsection}{}
\makeatother

\makeatletter
\@addtoreset{equation}{section}
\@addtoreset{figure}{section}
\@addtoreset{table}{section}
\makeatother

\usepackage{color}
\definecolor{bleudefrance}{rgb}{0.19, 0.55, 0.91}
\definecolor{royalblue}{rgb}{0.25, 0.41, 0.88}

\usepackage{feynmp-auto}

\usepackage{siunitx, booktabs}
\usepackage{polski} 

\usepackage{bibunits} 

%% file: pub101-orcid.tex
  \author{M.~Abumusabh\,\orcidlink{0009-0004-1031-5425}} 
  \author{I.~Adachi\,\orcidlink{0000-0003-2287-0173}} 
  \author{K.~Adamczyk\,\orcidlink{0000-0001-6208-0876}} 
  \author{A.~Aggarwal\,\orcidlink{0000-0002-5623-3896}} 
  \author{L.~Aggarwal\,\orcidlink{0000-0002-0909-7537}} 
  \author{H.~Ahmed\,\orcidlink{0000-0003-3976-7498}} 
  \author{Y.~Ahn\,\orcidlink{0000-0001-6820-0576}} 
  \author{H.~Aihara\,\orcidlink{0000-0002-1907-5964}} 
  \author{M.~Akdag\,\orcidlink{0009-0004-3728-1077}} 
  \author{N.~Akopov\,\orcidlink{0000-0002-4425-2096}} 
% ...
\collaboration{The Belle II Collaboration}

%% file: acknowledgements-b2.tex
This work, based on data collected using the Belle II detector, which was built and commissioned prior to March 2019,
was supported by
Higher Education and Science Committee of the Republic of Armenia Grant No.~23LCG-1C011;
Australian Research Council and Research Grants
No.~DP200101792, 
No.~DP210101900, 
No.~DP210102831, 
No.~DE220100462, 
No.~LE210100098, 
and
No.~LE230100085; 
Austrian Federal Ministry of Education, Science and Research,
Austrian Science Fund (FWF) Grants
DOI:~10.55776/P34529,
DOI:~10.55776/J4731,
DOI:~10.55776/J4625,
DOI:~10.55776/M3153,
and
DOI:~10.55776/PAT1836324,
and
Horizon 2020 ERC Starting Grant No.~947006 ``InterLeptons'';
Natural Sciences and Engineering Research Council of Canada, Digital Research Alliance of Canada, and Canada Foundation for Innovation;
National Key R\&D Program of China under Contract No.~2024YFA1610503,
and
No.~2024YFA1610504
National Natural Science Foundation of China and Research Grants
No.~11575017,
No.~11761141009,
No.~11705209,
No.~11975076,
No.~12135005,
No.~12150004,
No.~12161141008,
No.~12405099,
No.~12475093,
and
No.~12175041,
and Shandong Provincial Natural Science Foundation Project~ZR2022JQ02;
the Czech Science Foundation Grant No. 22-18469S,  Regional funds of EU/MEYS: OPJAK
FORTE CZ.02.01.01/00/22\_008/0004632 
and
Charles University Grant Agency project No. 246122;
European Research Council, Seventh Framework PIEF-GA-2013-622527,
Horizon 2020 ERC-Advanced Grants No.~267104 and No.~884719,
Horizon 2020 ERC-Consolidator Grant No.~819127,
Horizon 2020 Marie Sklodowska-Curie Grant Agreement No.~700525 ``NIOBE''
and
No.~101026516,
and
Horizon Europe Marie Sklodowska-Curie Staff Exchange project JENNIFER3 Grant Agreement No.~101183137 (European grants);
L’Institut National de Physique Nucl\'eaire et de Physique des
Particules (IN2P3) du CNRS under Project Identification No.
CNRS-IN2P3-14-PP-033
and L’Agence Nationale de la Recherche (ANR) under Grant No. ANR-23-CE31-
0018 and ANR-25-CE31-1333 (France);
BMFTR, DFG, HGF, MPG, and AvH Foundation (Germany);
Department of Atomic Energy under Project Identification No.~RTI 4002,
Department of Science and Technology,
and
UPES SEED funding programs
No.~UPES/R\&D-SEED-INFRA/17052023/01 and
No.~UPES/R\&D-SOE/20062022/06 (India);
Israel Science Foundation Grant No.~2476/17,
U.S.-Israel Binational Science Foundation Grant No.~2016113, and
Israel Ministry of Science Grant No.~3-16543;
Istituto Nazionale di Fisica Nucleare and the Research Grants BELLE2,
and
the ICSC – Centro Nazionale di Ricerca in High Performance Computing, Big Data and Quantum Computing, funded by European Union – NextGenerationEU;
Japan Society for the Promotion of Science, Grant-in-Aid for Scientific Research Grants
No.~16H03993,
No.~16H06492,
No.~16K05323,
No.~17H01133,
No.~17H05405,
No.~18K03621,
No.~18H03710,
No.~18H05226,
No.~19H00682, 
No.~20H05850,
No.~20H05858,
No.~22H00144,
No.~22K14056,
No.~22K21347,
No.~23H05433,
No.~26220706,
No.~26400255,
and
No.~26H02056,
and
the Ministry of Education, Culture, Sports, Science, and Technology (MEXT) of Japan;  
National Research Foundation (NRF) of Korea Grants
No.~2021R1-F1A-1064008,
No.~2022R1-A2C-1003993,
No.~RS-2018-NR031074,
No.~RS-2021-NR060129,
No.~RS-2024-00354342,
No.~RS-2025-02219521,
No.~RS-2026-25471491,
No.~RS-2026-25480677,
and
No.~RS-2026-25486791,
Radiation Science Research Institute,
Foreign Large-Size Research Facility Application Supporting project,
the Global Science Experimental Data Hub Center, the Korea Institute of Science and
Technology Information (K26L1M2C3)
and
KREONET/GLORIAD;
Universiti Malaya RU grant, Akademi Sains Malaysia, and Ministry of Education Malaysia;
Frontiers of Science Program Contracts
No.~FOINS-296,
No.~CB-221329,
No.~CB-236394,
No.~CB-254409,
and
No.~CB-180023, and SEP-CINVESTAV Research Grant No.~237 (Mexico);
the Polish Ministry of Science and Higher Education and the National Science Center;
the Ministry of Science and Higher Education of the Russian Federation
and
the HSE University Basic Research Program, Moscow;
University of Tabuk Research Grants
No.~S-0256-1438 and No.~S-0280-1439 (Saudi Arabia);
Slovenian Research Agency and Research Grants
No.~J1-50010
and
No.~P1-0135;
Ikerbasque, Basque Foundation for Science,
Basque Government through grant IT1977-26, State Agency for Research of the Spanish Ministry of Science and Innovation through Grants No. PID2022-136510NB-C33 and PID2025-174203NB-I00, Spain,
the Severo Ochoa project CEX2023-001292-S funded by MICIU/AEI, State Secretariat for
Telecommunications and Digital Infrastructure with reference
TSI-069100-2023-0012, State Agency for Research of the Spanish Ministry
of Science, Innovation and Universities through Grant No
PID2024-156645NB-C21;
The Knut and Alice Wallenberg Foundation (Sweden), Contracts No.~2021.0174, No.~2021.0299, and No.~2023.0315;
National Science and Technology Council,
and
Ministry of Education (Taiwan);
Thailand Center of Excellence in Physics;
TUBITAK ULAKBIM (Turkey);
National Research Foundation of Ukraine, Project No.~2020.02/0257,
and
Ministry of Education and Science of Ukraine;
the U.S. National Science Foundation and Research Grants
No.~PHY-1913789 
and
No.~PHY-2111604, 
and the U.S. Department of Energy and Research Awards
No.~DE-AC06-76RLO1830, 
No.~DE-SC0007983, 
No.~DE-SC0009824, 
No.~DE-SC0009973, 
No.~DE-SC0010007, 
No.~DE-SC0010073, 
No.~DE-SC0010118, 
No.~DE-SC0010504, 
No.~DE-SC0011784, 
No.~DE-SC0012704, 
No.~DE-SC0019230, 
No.~DE-SC0021616, 
No.~DE-SC0022350, 
No.~DE-SC0023470; 
and
the Vietnam Academy of Science and Technology (VAST) under Grant
No.~DL0000.05/26-27.

These acknowledgements are not to be interpreted as an endorsement of any statement made
by any of our institutes, funding agencies, governments, or their representatives.

We thank the SuperKEKB team for delivering high-luminosity collisions;
the KEK cryogenics group for the efficient operation of the detector solenoid magnet and IBBelle on site;
the KEK Computer Research Center for on-site computing support; the NII for SINET6 network support;
and the raw-data centers hosted by BNL, DESY, GridKa, IN2P3, INFN, 
and the University of Victoria.